\documentclass[aps,prl,twocolumn,showpacs]{revtex4}
\usepackage{graphicx}
\usepackage{dcolumn}
\setcitestyle{super}

\begin{document}

\preprint{APS/123-QED}

\title{Resistivity and magnetoresistance of FeSe single crystals under Helium-gas pressure}
\author{S. Kn\"{o}ner$^{1}$}
\author{D. Zielke$^{1}$}
\author{S. K\"{o}hler$^{1}$}
\author{B. Wolf$^{1}$}
\author{Th. Wolf$^{2}$}
\author{L. Wang$^{2}$}
\author{A. B\"{o}hmer$^{2,3}$}
\author{C. Meingast$^{2}$}
\author{M. Lang$^{1}$}

\address{$^{1}$Physikalisches Institut, J.W. Goethe-Universit\"{a}t
Frankfurt(M), SPP1458, D-60438 Frankfurt(M), Germany}
\address{$^{2}$Institute for Solid State Physics, Karlsruhe Institute of Technology, D-76131 Karlsruhe, Germany}
\address{$^{3}$Ames Laboratory, U.S. DOE, Iowa and Department of Physics and Astronomy, Iowa State University, Ames, Iowa 50011, USA}

\date{\today}

\begin{abstract}
We present temperature-dependent in-plane resistivity measurements on FeSe single crystals under He-gas pressure up to 800\,MPa and magnetic fields $B \leq$ 10\,T. A sharp phase transition anomaly is revealed at the tetragonal-to-orthorhombic transition at $T_s$ slightly below 90\,K. $T_s$ becomes reduced with increasing pressure in a linear fashion at a rate  d$T_{s}$/d$P$ $\simeq$ -31\,K/GPa. This is accompanied by a $P$-linear increase of the superconducting transition temperature at $T_c \sim$ 8.6\,K with d$T_{c}$/d$P$ $\simeq$ +5.8\,K/GPa.
Pressure studies of the normal-state resistivity highlight two distinctly different regimes: for $T > T_s$, i.e., in the tetragonal phase, the in-plane resistivity changes strongly with pressure. This contrasts with the state deep in the orthorhombic phase at $T \ll T_s$, preceding the superconducting transition. Here a $T$-linear resistivity is observed the slope of which does not change with pressure. Resistivity studies in varying magnetic fields both at ambient and finite pressure reveal clear changes of the magnetoresistance, $\Delta \rho \propto B^{2}$, upon cooling through $T_s$. Our data are consistent with a reconstruction of the Fermi surface accompanying the structural transition.

\end{abstract}


\maketitle
\section{I. INTRODUCTION}
The discovery of superconductivity in LaFeAsO \cite{Hosono2008} and in related iron-pnictide or -chalcogenide families \cite{Hsu2008, Wang2008, Rotter2008a} has attracted enormous interest to this class of materials. The parent compounds of these so-called 1111 or 122 systems are typically metallic paramagnets which undergo a tetragonal/paramagnetic to orthorhombic/antiferromagnetic phase transition upon cooling. The magnetic transition at $T_N$ often occurs at the same or a slightly lower temperature than the structural transition at $T_s$. Both $T_N$ and $T_s$ can be suppressed by the application of pressure or by chemical substitution giving way to superconductivity \cite{Paglione2010, Johnston2010, Steward2011}. The mutual relation between structure, magnetism and superconductivity remains a key issue in this class of superconductors. Open questions are whether or not the structural transition is driven by the magnetic degrees of freedom, or vice versa, and whether magnetic or orbital fluctuations are responsible for the superconducting pairing \cite{Fernandes2012, Kontani2011, fernandes3}. The two types of order are closely related by symmetry, both breaking the $C4$ rotational symmetry of the tetragonal/paramagnetic phase, which can be associated with a so-called nematic phase \cite{Chu2010, Chuang2010, Fernandes2010}.\\ 
The iron-based superconductors all share a common structural feature, i.e., iron arsenic/selenium layers which are separated by intermediate layers. Among the various iron-arsenide and -chalcogenide families, FeSe is of particular interest because of its simple structure without spacing layers \cite{Paglione2010, buechner}. Moreover, FeSe shows a structural phase transition at $T_{s}$ $\sim$ 90 K, similar to that found in the related 1111- and 122-parent compounds \cite{Johnston2010}, but, contrary to the latter systems, shows magnetic order only at high pressure \cite{Bendele2010, McQueen2010, Terashima2015}. Superconductivity, emerging in FeSe around 8\,K out of an orthorhombic structure \cite{Hsu2008}, can be significantly enhanced up to $T_c$ values around 37\,K through the application of pressure \cite{Medvedev2009, Masaki2009, Bendele2010} which drives the system tetragonal \cite{Masaki2009}. Based on recent angle-resolved photoemission \cite{Shimojima2014, Nakayama2014} and NMR \cite{Baek2014, Boehmer2015} studies it has been proposed that, in contrast to the 122 family, where magnetic degrees of freedom have been discussed as the origin for nematicity \cite{fernandes3, Fernandes2010}, for FeSe $T_s$ is due to orbital order. These observations together with recent theoretical proposals on frustrated magnetism \cite{Wang2015, Glasbrenner2015} make FeSe a particularly interesting case for investigating the relation of structural and magnetic degrees of freedom and their role for superconductivity. 
In this work, we use resistivity measurements under varying magnetic field and hydrostatic ($^4$He-gas) pressure for studying the various phases and phase transitions in FeSe. \\

\section{II. EXPERIMENTAL DETAILS}

High-quality single crystals of FeSe were synthesized by employing a vapor-transport technique as described elsewhere \cite{Boehmer2013}. Measurements were performed on three crystals with approximate dimensions of 2$\times$2$\times$0.2\,mm$^{3}$ (sample \#1), 1$\times$1$\times$0.05\,mm$^{3}$ (sample \#2), and 3.2$\times$1.2$\times$0.15\,mm$^{3}$ (sample \#3). Sample \#2 was obtained by cleaving sample \#1 after completion of a first set of experiments. The electrical resistivity was measured by using a standard four-terminal $dc$-technique with an applied current of $I$ = 1\,mA. Four 25\,$\mu$m gold wires were fixed with silver paste in the $ab$ plane. Data were taken both as a function of temperature upon cooling the sample at a rate of -0.2\,K/min or as a function of magnetic field $B \leq$ 10\,T at constant temperature. Measurements in magnetic fields were performed with $B$ aligned parallel to the $ab$ plane and along the $c$ direction, respectively, and the current $I$ $\bot$ $B$. For measurements under pressure $P \leq$ 800\,MPa a CuBe pressure cell (Institute of High-Pressure Physics, Polish Academy of Sciences, Unipress Equipment Division) was used with $^{4}$He as a pressure-transmitting medium. The cell is connected by a CuBe capillary to a He-gas compressor kept at room temperature to ensure that $P \simeq$ const. during $T$ sweeps. An n-InSb \cite{Kraak1984} single crystal was used for an \textit{in situ} determination of $P$. The use of helium as a pressure-transmitting medium ensures truly hydrostatic pressure conditions as long as it is in the liquid phase, i.e., in a $T$-$P$ range above the solidification line $T_{He}^{solid}$($P$), see below. Even when cooling through $T_{He}^{solid}$($P$), which is accompanied by a pressure loss of about 30\%, c.f.\,the inset to Fig.\,\ref{fig:2}, deviations from hydrostatic conditions are small. This is due to the low solidification temperature of helium, implying a small thermal expansion mismatch between sample and frozen pressure medium, and the small shear modulus of solid helium \cite{Smhe4}. As has been demonstrated by various groups, hydrostaticity can be an important issue in studying the properties of iron-based superconductors \cite{Yu2009, Canfield2009, Gati2013, Tafour2014}.\\
\section{III. RESULTS}
\subsection{A. Electrical resistivity under hydrostatic pressure}
\begin{figure}
	\centering
		\includegraphics[width=1.00\columnwidth]{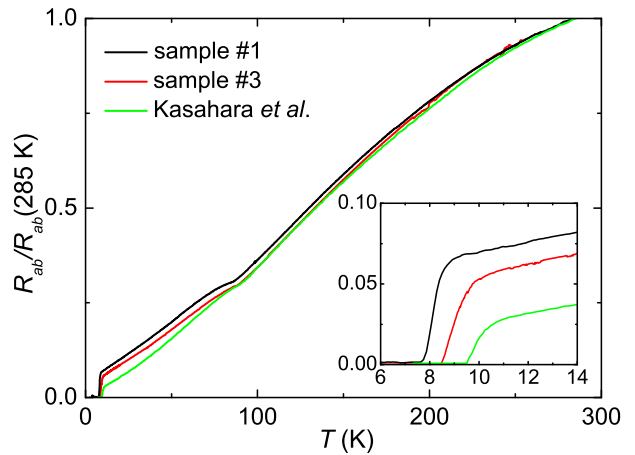}
		\caption{Normalized in-plane resistance of three FeSe single crystals (\#1 and \#3 studied in this work and a crystal investigated by Kasahara \textit{et al.} \cite{Kasahara2014}) from the same source (Karlsruhe Institute of Technology, KIT). The inset shows the data around the superconducting transition on enlarged scales.}
	\label{fig:1}
\end{figure}

Figure \ref{fig:1} shows the in-plane resistance data, normalized to their value at 285\,K, for the single crystals \#1 and \#3 as a function of temperature. For comparison, the figure also includes data taken on another FeSe crystal from the same source previously studied by Kasahara \textit{et al.} \cite{Kasahara2014}. For all three crystals the resistance shows a sharp kink at temperatures 86.4\,K (\#1), 84.5\,K (\#2) and 89.3\,K (\#3) which is assigned to the structural transition at $T_s$. Upon cooling through $T_{s}$ the resistance exhibits a well-defined change of slope at $T_{s}$, such that the resistivity is larger in the low-temperature phase than the extrapolated value of the high-temperature phase. We note that this contrasts with the behavior observed at the usual SDW transition in, e.g., clean Ba-122 or 1111 compounds \cite{Hosono2008, Rotter2008a, Klauss2008} where the resistivity sharply decreases below $T_{s}$. While the data sets for the three crystals almost collapse for $T > T_s$, there is a considerable sample-to-sample variation in $R_{ab}$($T$) for temperatures below $T_{s}$. This might indicate different residual resistivity ratios RRR $\simeq$ $R$(285\,K)/$R$(10\,K), and thus different sample qualities. However, since cooling through $T_s$ is accompanied by twinning, different twin states for these crystals together with a finite in-plane anisotropy in the resistivity ($\rho_a$ vs. $\rho_b$) may also contribute to the different behaviors of $R_{ab}$($T$) for $T < T_s$. 

\begin{figure}
	\centering
		\includegraphics[width=1.00\columnwidth]{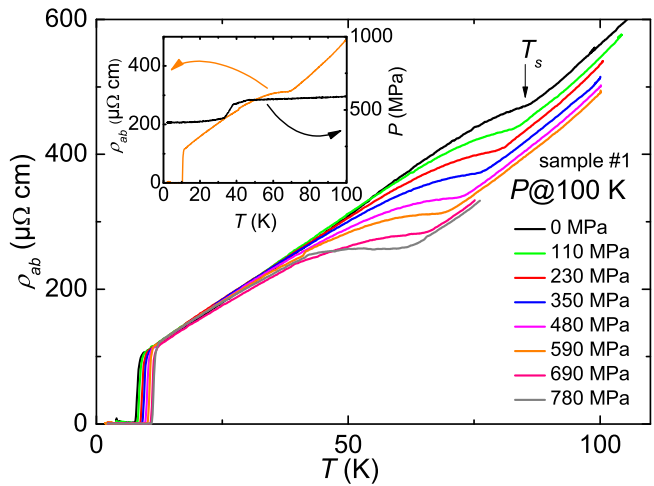}
		\caption{Electrical resistivity of single crystalline FeSe (sample \#1) at varying pressures up to $P$ = 780\,MPa applied at 100\,K. The inset shows the evolution of the pressure, initially set to $P$ = 590\,MPa, upon cooling. An abrupt pressure loss of about 30\% occurs due to the solidification of the helium at 46 K for this pressure value.}
	\label{fig:2}
\end{figure}

\begin{figure}
	\centering
		\includegraphics[width=1.00\columnwidth]{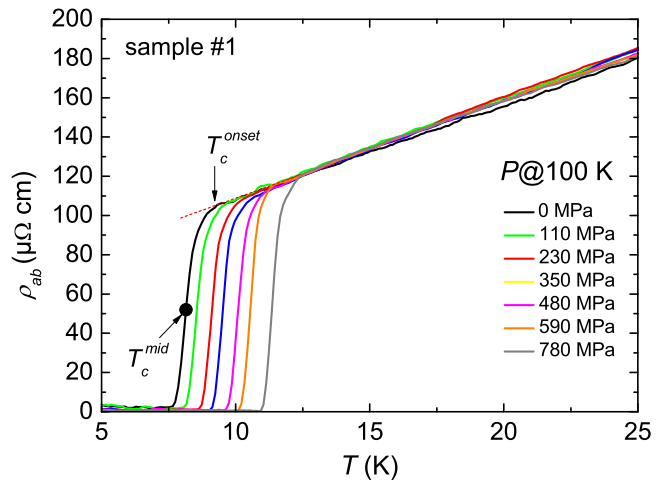}
		\caption{In-plane resistivity data for single crystalline FeSe slightly above and around the superconducting transition temperature under varying hydrostatic pressure values up to 780\,MPa. The onset of superconductivity, $T^{onset}_c$, is defined as the point where the data deviate form the in-$T$ linear low-temperature behavior, and $T^{mid}_c$ where the drop amounts to 50\% of the extrapolated linear behavior.}
	\label{fig:3}
\end{figure}

\begin{figure}
	\centering
		\includegraphics[width=1.00\columnwidth]{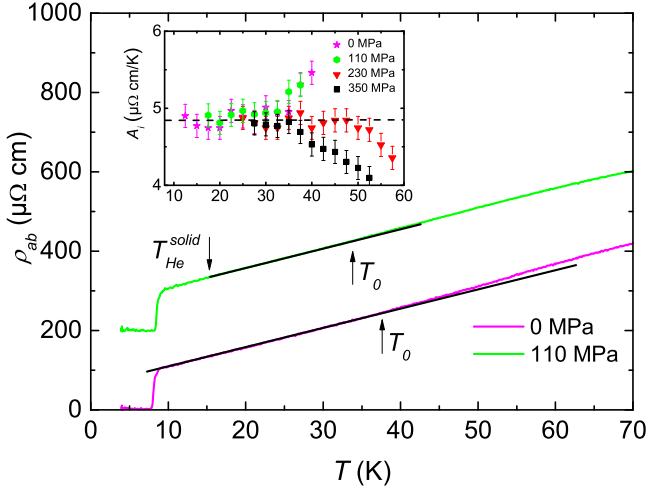}
		\caption{In-plane resistivity data for sample  \#1 at $P$ = 0 and 110\,MPa for $T \leq$ 70\,K. $T_0$ (up arrow) marks the temperature above which $A_i$($T$) departs from the average slope (represented by the black solid line) by one standard deviation. $T_{He}^{solid}$ (down arrow) indicates the solidification temperature of $^{4}$He. The data for $P$ = 110\,MPa were shifted vertically for clarity. The inset shows the temperature variation of the slopes $A_i$, determined as described in the text, for $P$ = 0, 110, 230, and 350\,MPa.}
	\label{fig:4}
\end{figure}

In the inset of Fig.\,\ref{fig:1} we show the data around the superconducting transition. We find $T_c$ values (midpoints of the transition) of 8\,K for sample \#1, 9\,K for sample \#3 and 10\,K for the crystal studied by Kasahara \textit{et al.} \cite{Kasahara2014}. The data also indicate differences in the transition width $\Delta T_c$ (90\% - 10\% of the normal state resistance) ranging between 0.85 K (\#1), 2.2 K (\#2) and  1.15 K (\#3).\\
Figure \ref{fig:2} shows the in-plane resistivity $\rho_{ab}$ of sample \#1 for various pressures up to 780\,MPa. The pressure was applied at 100\,K, i.e., in the liquid phase of the pressure-transmitting medium helium. The following basic observations were made: (1) the structural transition temperature $T_{s}$ becomes significantly suppressed with increasing the pressure. (2) There is a considerable pressure-induced reduction of the in-plane resistivity for $T > T_s$, i.e., in the tetragonal phase. This contrasts with a much weaker effect of hydrostatic pressure on the low-temperature normal state at $T \ll T_s$($P$), i.e., in the orthorhombic phase, preceding the transition in the superconducting state. For a closer inspection of the latter aspect and for highlighting the effect of pressure on $T_c$, we show in Figs.\,\ref{fig:3} and \ref{fig:4} a selection of low-temperature data on expanded scales. Figure \ref{fig:3} demonstrates that (3) $T_c$ is monotonously raised with increasing pressure in the entire pressure range $P \leq$ 780\,MPa while the transition width remains unaffected. Moreover, as will be explicated below in more detail, the data reveal (4) an extended $T$-linear range the slope of which does not change with pressure.\\
For analyzing the temperature dependence of the normal-state resistivity shown in Fig.\,\ref{fig:2} for $T \ll T_s$($P$), the following procedure was applied: the temperature range of interest was divided into intervals $\Delta T_i$ = $T_{i+1}$ - $T_i$ = 5\,K ($T_1$ = 10\,K), in each of which the data were fitted by a linear function $\rho_{ab}^{i}$($T$) = $\rho_0^{i}$ + $A_i T$. The variation of the slope $A_i$ with temperature is shown in the inset of Fig.\,\ref{fig:4}. For the data at $P$ = 0, $A_i$ stays constant within the error bars at $A$ = (4.9 $\pm$ 0.15) $\mu \Omega$ cm for 10\,K $\leq T \leq T_0$ $\simeq$ 35\,K, i.e., $\rho$($T$) follows to a good approximation an in-$T$ linear behavior (solid line in Fig.\,\ref{fig:4}). $T_0$ marks the temperature above which $A_i$ departs from the average slope by one standard deviation. The same procedure was applied also to the data at finite pressure, exemplarily shown for $P$ = 110\,MPa in Fig.\,\ref{fig:4}. For these data, we constrained the fits to temperatures $T > T_{He}^{solid}$($P$) (indicated by a down arrow in Fig.\,\ref{fig:4}) to rule out artefacts resulting from the solidification process of helium. As shown in the inset of Fig.\,\ref{fig:4}, the average $A_i$ remains virtually unchanged upon increasing the pressure to $P$ = 110\,MPa, $P$ = 230\,MPa and $P$ = 350\,MPa. For the latter pressure values, $A_i$ falls below the average value at higher temperatures $T$ $\gtrsim$ 35 K. We assign this effect to the nearby structural transition which gives rise to a reduced slope d$\rho$/d$T$ over a considerable temperature window $T < T_s$, cf.\,Fig.\,\ref{fig:2}. Hence, the seemingly enhanced value of $T_0$ for $P$ = 230\,MPa might be an artefact resulting from the compensation of two counteracting effects: the upward deviations from the average $A_i$ for $T > T_0$ revealed at low pressures (c.f. the data at $P$ = 0 and 110\,MPa in the inset of Fig.\,\ref{fig:4}) is likely to be partly compensated by the downward deviation caused by the structural transition at this pressure value. Due to the growing influence of effects related to $T_s$ with increasing pressure and the simultaneous increase of $T_{He}^{solid}$($P$), we refrain from analyzing the data at pressures $P >$ 350\,MPa.


\begin{figure}
	\centering
		\includegraphics[width=1.00\columnwidth]{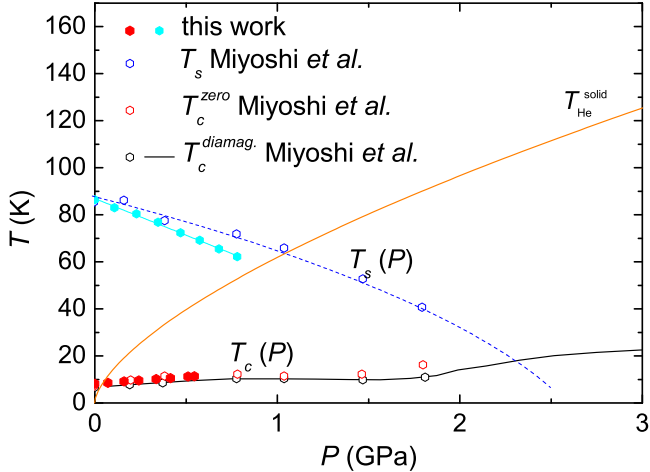}
		\caption{Compilation of the transition temperatures $T_s$($P$) and $T_c$($P$) in a temperature-pressure phase diagram for FeSe derived from this work (filled symbols). $T^{solid}_{He}$ (orange solid line) marks the solidification line of helium used as a pressure-transmitting medium. For those pressures where $T_c$($P$) lies below $T^{solid}_{He}$($P$), the corresponding pressure values have been corrected for the 30\% pressure loss accompanying the solidification of helium, cf.\,the inset of figure \ref{fig:2}. Also shown are the pressure data reported by Miyoshi \textit{et al.} \cite{Miyoshi2014} taken on single crystalline FeSe employing oil as a pressure-transmitting medium: $T_s$($P$) (blue open circles) and $T_{c}^{zero}$ (red open circles) were obtained via transport measurements, $T_{c}^{diamag}$ (black open symbols and black solid line) corresponds to the onset of diamagnetism in susceptibility measurements.} 
	\label{fig:5}
\end{figure}

Figure \ref{fig:5} compiles the transition temperatures $T_s$($P$) and $T_c$($P$) in a $T$-$P$ phase diagram. For the structural transition at $T_s$ the data taken on sample \#1 reveal to a good approximation a linear decrease with a slope d$T_{s}$/d$P$ = -(31.1 $\pm$ 0.7)\,K/GPa. A linear variation with pressure is also found for the superconducting transition temperature. Using the midpoint, $T_c^{mid}$, or the onset temperature, $T_c^{onset}$, both are defined in Fig.\,\ref{fig:3}, yields an increase of $T_c$ with pressure with a rate d$T_{c}$/d$P$ = +(5.84 $\pm$ 0.06)\,K/GPa. For comparison Fig.\,\ref{fig:5} also includes pressure data on $T_s$ and $T_c$ reported by Miyoshi \textit{et al.} \cite{Miyoshi2014} by using oil as pressure-transmitting medium.  

\subsection{B. Magnetoresistance effects at zero and finite hydrostatic pressure}

\begin{figure}
	\centering
		\includegraphics[width=1.00\columnwidth]{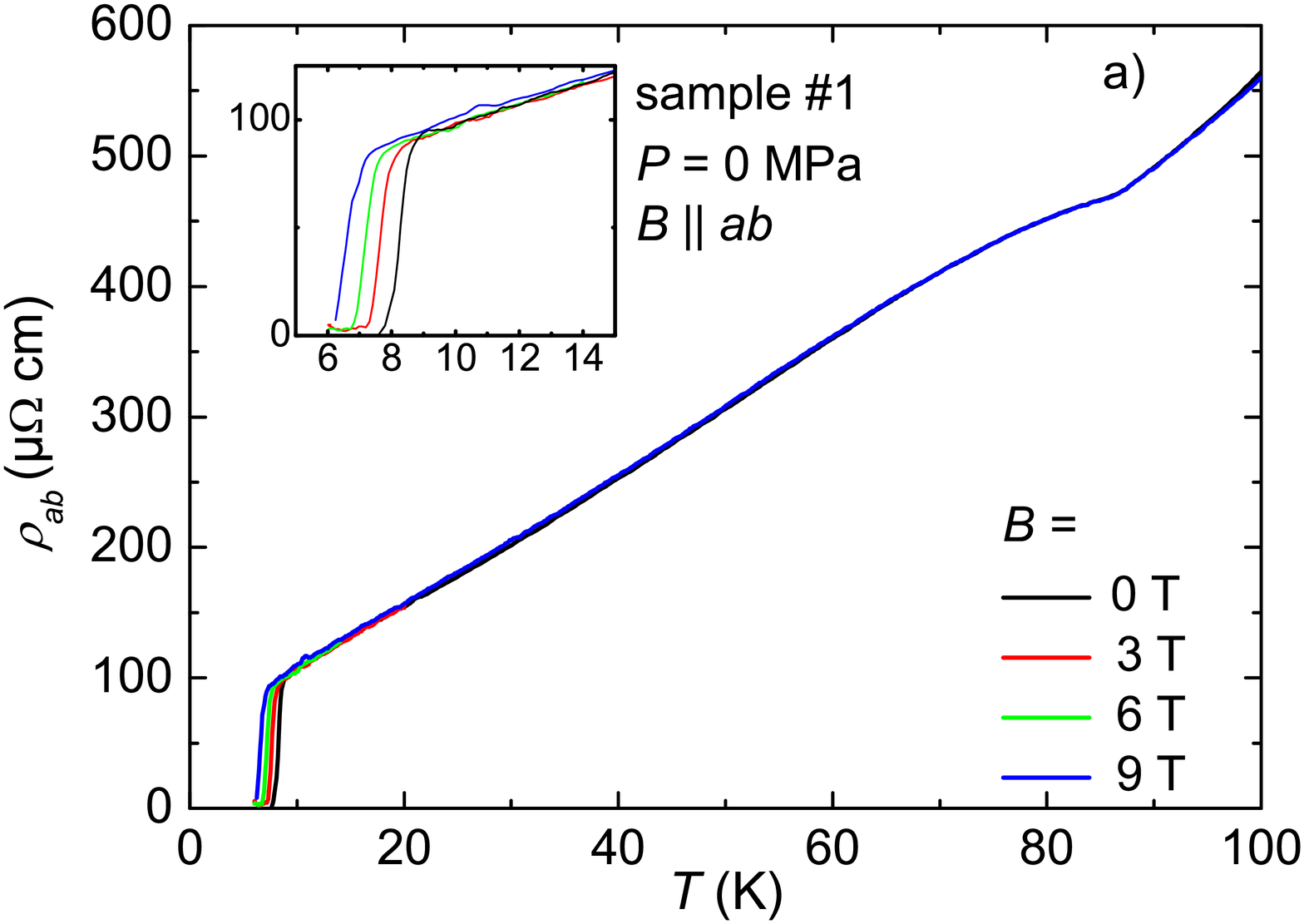}
		\includegraphics[width=1.00\columnwidth]{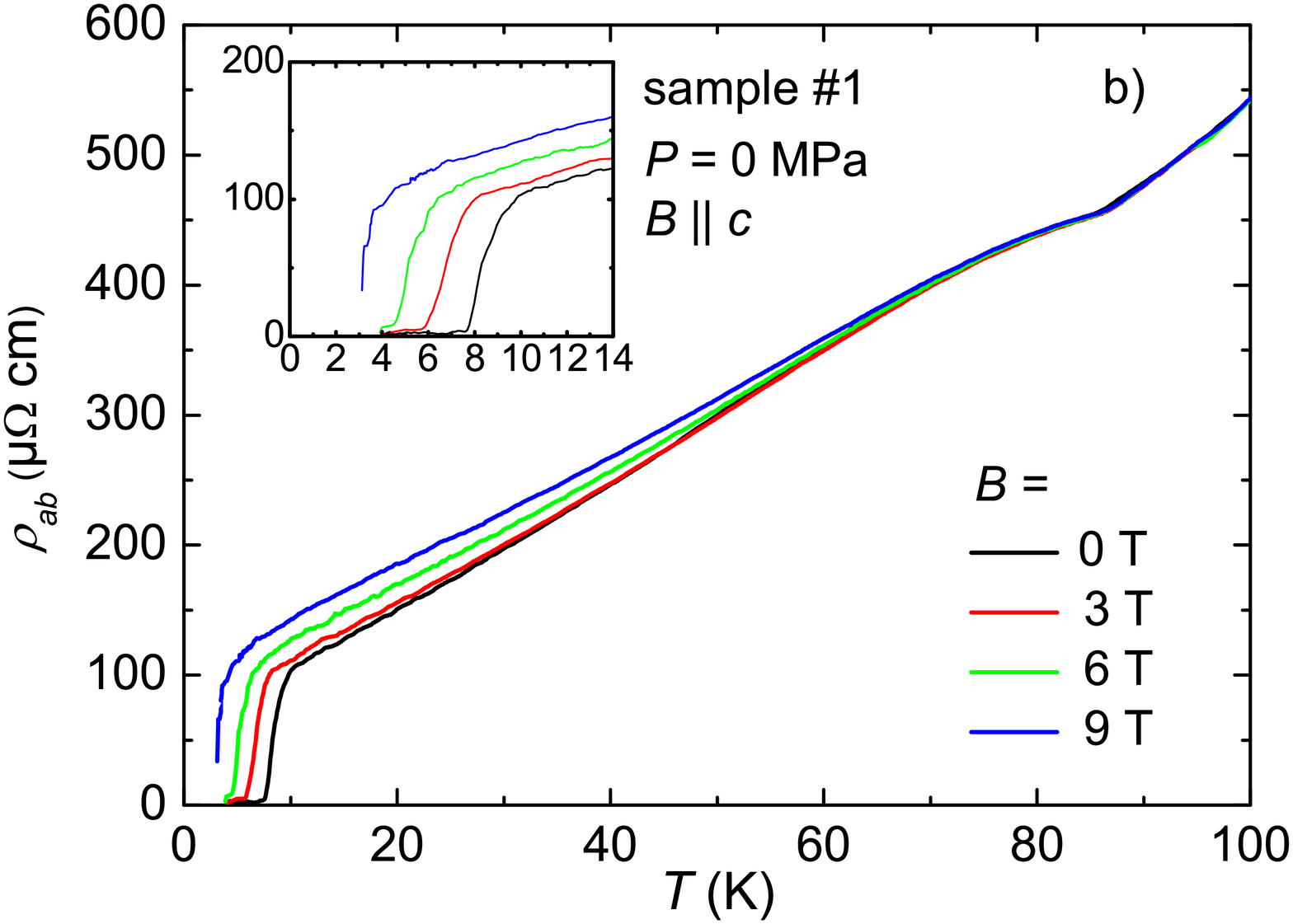}
		\caption {In-plane resistivity of single crystalline FeSe (sample \#1) at ambient pressure for magnetic fields $B \leq$ 9\,T aligned parallel to the $ab$ plane (a) and parallel to the $c$ axis (b). The insets show the data in the vicinity of the superconducting transition on expanded scales.}
	\label{fig:6}
\end{figure}

In Fig.\,\ref{fig:6} we show the in-plane resistivity at ambient pressure in varying magnetic fields $B$ aligned parallel to the $ab$ plane (Fig.\,\ref{fig:6}a) and parallel to the $c$ axis 
(Fig.\,\ref{fig:6}b). The insets of Fig.\,\ref{fig:6}a and \ref{fig:6}b show details around the superconducting transition. For fields parallel to the $ab$-plane (Fig.\,\ref{fig:6}a) up to 9\,T, the maximum field used in this experiment, we observe a reduction of $T_{c}$ by about -(0.185 $\pm$ 0.01)\,K/T. Apart from the shift in $T_c$ there is, within the resolution of our experiment, no other noticeable effect of a field $B$ aligned along the $ab$ on the normal-state resistivity for $T \leq$ 100\,K.

For magnetic fields aligned parallel to the $c$ axis (Fig.\,\ref{fig:6}b), $T_c$ becomes suppressed at a distinctly stronger rate of -(0.56 $\pm$ 0.02)\,K/T, cf.\,inset to Fig.\,\ref{fig:6}b. A similar anisotropy in the initial slopes of the upper critical fields has been reported by Terashima \textit{et al.} \cite{Terashima2014}. 

In addition, the data in Fig.\,\ref{fig:6}b uncover a finite magnetoresistance (MR) effect over a wide range of temperature which grows upon cooling as observed previously \cite{Kasahara2014, Huynh2014, Watson2015a}. 

\begin{figure}
	\centering
		\includegraphics[width=1.00\columnwidth]{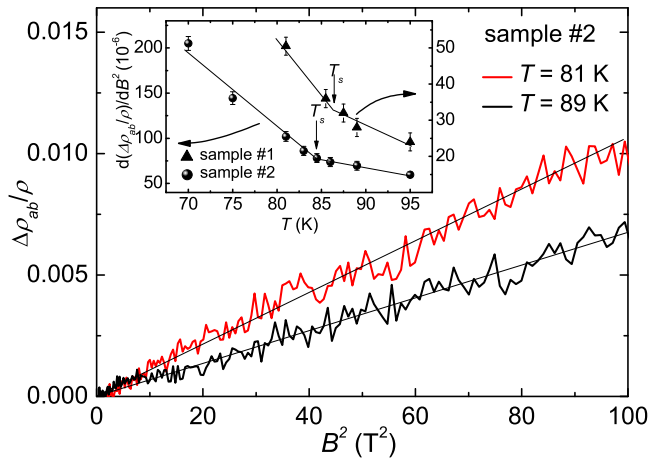}
		\caption{Normalized in-plane magnetoresistance $\Delta \rho_{ab}$ = $\rho_{ab}(B)$-$\rho_{ab}(0)$ measured at temperatures slightly below (red solid line for $T_1$ = 81\,K) and above (black solid line for $T_2$ = 89\,K) the structural transition at $T_s$ = 86.5\,K of sample \#2 plotted vs. $B^{2}$. Thin straight lines are guides to the eyes. The inset shows the slopes d($\Delta \rho_{ab}$/$\rho_{ab}$)/d$B^{2}$ vs. temperature for sample \#1 (filled triangles) and sample \#2 (filled circles) derived from isothermal field sweeps as shown in the main panel. The arrows at $T_s$ mark the structural transition temperatures as determined from the kinks in the resistivity.}
	\label{fig:7}
\end{figure}

For a more detailed investigation of the MR effect, especially for exploring its interrelation with the structural transition at $T_s$, experiments at $T$ = const. under varying magnetic fields have been performed. In the main part of Fig.\,\ref{fig:7} we show results of $\frac{\Delta\rho_{ab}}{\rho_{ab}(0)}=\frac{\rho_{ab}(B)-\rho_{ab}(0)}{\rho_{ab}(0)}$ for two fixed temperatures, $T_1$ = 81\,K and $T_2$ = 89\,K, i.e., slightly below and above $T_s$, respectively. In these experiments the temperature was stabilized within $\Delta T$ = $\pm$ 0.01\,K. For both temperatures, the MR $\Delta \rho_{ab}$($B$) for $B \leq$ 10\,T, the maximum field available, follows to a good approximation a $B^{2}$ dependence albeit with different slopes. A $\Delta \rho_{ab}$($B$) $\propto B^{2}$ dependence was observed also for all other $T$ = const. experiments on samples \#1 and \#2 performed here. The slopes, d($\Delta \rho_{ab}$)/d$B^{2}$ for samples \#1 and \#2 are plotted as a function of temperature in the inset of Fig.\,\ref{fig:7}. The figure demonstrates that cooling through $T_s$ is accompanied by an abrupt increase in the slope, i.e., a change in the MR effect.

\begin{figure}
	\centering
		\includegraphics[width=1.00\columnwidth]{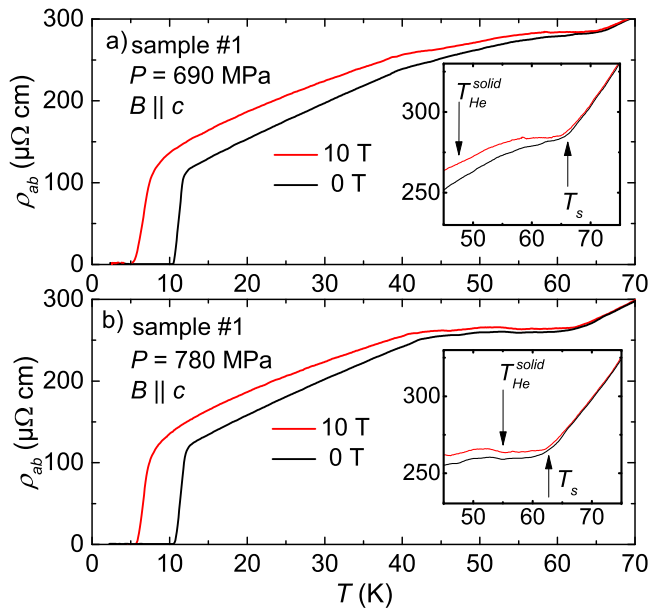}
		\caption{Temperature dependence of the in-plane resistivity for single crystalline FeSe sample \#1 under hydrostatic pressure of 690\,MPa (a) and 780\,MPa (b) measured in $B$ = 0 (black lines) and 10\,T (red lines) applied parallel to the $c$ axis. The insets show the data around the structural transition $T_s$ marked by an up arrow. The down arrows indicate the solidification temperature at the given pressure value.}
	\label{fig:8}
\end{figure}

To follow the MR effect and its interrelation with $T_s$ also under varying hydrostatic pressures, we compare in Fig.\,\ref{fig:8} temperature runs taken at zero field and at $B$ = 10\,T, applied parallel to the $c$ axis, for pressure values of $P$ = 690\,MPa (Fig.\,\ref{fig:8}a) and $P$ = 780\,MPa (Fig.\,\ref{fig:8}b). For these experiments $T$ sweeps at different constant fields are preferred over $B$ sweeps due to difficulties in securing $T$ = const. conditions over a sufficiently long period of time under finite pressure. The data in the main panels and the insets of Fig.\,\ref{fig:8} demonstrate that the MR effect abruptly grows below $T_s$ also under finite pressure conditions.

\section{IV. DISCUSSION}

\subsection{A. Pressure dependences of $T_s$ and $T_c$}
The response we observed for both the structural and the superconducting transition of single-crystalline FeSe to hydrostatic pressure can be compared with results of other pressure studies on this material reported in the literature. The initial decrease in $T_s$ observed here of d$T_{s}$/d$P$ = -(31 $\pm$ 0.7)\,K/GPa is slightly larger than $\sim$ -22\,K/GPa revealed by Miyoshi \textit{et al.} \cite{Miyoshi2014} in their experiments where oil was used as a pressure-transmitting medium. On the other hand, for the superconducting transition, the initial slope of $T_c$ reported from these and other authors of 6-7 K/GPa \cite{Miyoshi2014, Bendele2010, Masaki2009} is very close to d$T_{c}$/d$P$ = +(5.8 $\pm$ 0.1)\,K/GPa observed here.

An independent determination of the initial slope of the pressure dependence for $T_c$ and $T_s$ can be performed by using results from thermodynamic measurements, carried out at ambient pressure. Since for FeSe also the structural transition is of second order, as demonstrated by thermal expansion measurements \cite{Boehmer2013}, the Ehrenfest relation can be applied for calculating the initial slope of both critical temperatures $T_{cr}$ = $T_s$ and $T_c$:

\begin{displaymath}
(\frac{dT_{cr}}{dp})_{p\rightarrow 0} = V_{m}\cdot T_c \frac{\Delta\beta}{\Delta C_{p}},
\end{displaymath}

where $V_m$ = 23.34 cm$^3$/mol is the molar volume, $\Delta\beta$ and $\Delta C_{p}$ are the corresponding jumps at $T_{cr}$ in the volume thermal expansion and specific heat, respectively. For the superconducting transition, by using $\Delta\beta$ = 2.14 $\cdot$ 10$^{-6}$ K$^{-1}$ and $\Delta C_{p}$/$T_{c}$ = 9.45 mJ mol$^{-1}$ K$^{-2}$\, \cite{lin2011}, this yields d$T_{c}$/d$P$ = (5.3 $\pm$ 1.2) K/GPa \cite{Boehmer2013}, in good agreement with the present results. For the structural transition, using $\Delta C_p/T_s$ $\approx$ 5.5 mJ mol$^{-1}$ K$^{-2}$\, \cite {Boehmer2015} and  $\Delta\beta$ $\approx$ 8.87 $\cdot$ 10$^{-6}$ K$^{-1}$\, \cite{Boehmer2013}, we find d$T_{s}$/d$P$ $\approx$ 37.6 K/GPa, also in good agreement with our present data. Whereas there appears to be a clear correlation between the increase of $T_{c}$ and the decrease of $T_{s}$ under pressure as found in other Fe-based materials \cite{Sefat2011}, which suggests a competition between these states, the effect of superconductivity upon the orthorombic distortion was found to be different in FeSe \cite{Boehmer2013}.

\subsection{B. Pressure dependences of the normal-state resistivity}

The data in Fig.\,\ref{fig:2} reveal a substantial decrease of the resistivity with increasing pressure in the tetragonal phase ($T > T_s$). At 100\,K the reduction is strictly linear in $P$ with a rate $R_{ab}^{-1} \Delta R_{ab}$(100\,K)/$\Delta P$ = -(22 $\pm$ 1)\%/GPa. A strong pressure effect was also observed for SrFe$_2$As$_2$ ($T_s$ = 200\,K) and BaFe$_2$As$_2$ ($T_s$ = 130\,K) in their high-temperature tetragonal phase where at 300\,K ($ > T_s$) a decrease of -9\%/GPa (SrFe$_2$As$_2$) and -(7-9)\%/GPa (BaFe$_2$As$_2$) was found \cite{Colombier2009, duncan2010}. Since FeSe has a much softer lattice than these 122 compounds, it is helpful to relate the response in the resistance to the associated relative volume change $\Delta V/V$, by using published data for the bulk modulus $B$ = -$V \Delta P/\Delta V$ of $\approx$ 30\,GPa for FeSe \cite{millican2009, margadonna2009} and 60 - 80\,GPa for the 122 systems \cite{kimber2009, uhoya2010, mittal2011, kasinathan2011}. For the quantity $R_{ab}^{-1}\Delta R_{ab}$(100\,K)/$\Delta V/V$  we find similar values of $\approx$ -660\% for FeSe and -(450-700)\% for SrFe$_2$As$_2$ and BaFe$_2$As$_2$. This indicates that the high sensisitivity of the tetragonal phase to pressure is a feature common to this class of compounds.

In contrast to the tetragonal phase, the resistivity in the orthorhombic phase upon approaching the superconducting state is found to be essentially pressure insensitive. For pressures $P \leq$ 350\,MPa an in-$T$ linear behavior $\rho_{ab}$($T$) = $\rho_{ab}$(0) + $A T^{\alpha}$, $\alpha$ = 1, with a pressure-independent slope $A$, could be identified over an extended temperature range. This is a remarkable observation in various respects. First, an $T$-linear resistivity over an extended temperature range has been observed also for other Fe-based superconductors \cite{Kasahara2010, doiron2009}. A particularly clear case has been revealed for the isovalently-substituted BaFe$_2$(As$_{1-x}$P$_x$)$_2$ \cite{Kasahara2010}. In this material, where the effect of P substitution for As can be directly linked to the effect of chemical pressure, an exponent $\alpha$ = 1 is seen only near optimum doping x = 0.33, i.e., where $T_c$ is maximum. This peculiar behavior in the resistivity and other transport properties has been assigned to antiferromagnetic fluctuations around the material's SDW quantum-critical point. This situation is markedly different from the present case for FeSe where the $T$-linear resistivity characterizes a state where $T_c$ is small, far below its maximum value. 

Furthermore, there is no obvious way to link the in-$T$ linear resistivity with the presence and the strength of antiferromagnetic spin fluctuations in FeSe observed by NMR experiments \cite{Imai2009}. According to these studies spin fluctuations in FeSe grow upon cooling below $T_s$ and become further enhanced by the application of pressure. Although the pressure values applied in ref.\,\cite{Imai2009} of 0.7 - 2.2\,GPa exceed those used here by more than a factor of 2, the absence of any measurable effect of pressure on the coefficient $A$ of the $T$-linear resistivity speaks against a common origin of both phenomena.

\subsection{C. Magnetoresistance and its interrelation with $T_s$}
From the temperature-dependent resistivity measurements performed under different field orientations (Fig.\,\ref{fig:6}) a finite transversal magnetoresistance (MR) effect $\Delta \rho_{ab} \propto B^{2}$ with no evidence of saturation was observed for $B \parallel c$. The small size of this effect, about 1\% at 10\,T, and its $B^{2}$ dependence are consistent with FeSe being a compensated semimetal \cite{Subedi2008,Singh2009,Kasahara2010,Huynh2014}, a notion which is further corroborated by recent high-field magnetotransport measurements \cite{Watson2015a}. For such systems, the simple relation $\Delta \rho$/$\rho$($0$) $\propto$ ($\omega_c \tau$) $^{2}$ may hold with $\omega_c$ the cyclotron frequency and $\tau$ the scattering time \cite{Pippard1989}. Thus, the absence of any measurable MR effect for $B \parallel ab$ may be the result of the altered orientation of the cyclotron orbits for this measuring geometry, now involving inter-plane contributions, together with a reduced inter-plane scattering time and/or an enhanced inter-plane effective carrier mass $m^{*}_{\bot} \propto \omega_c^{-1}$.\\
At the same time, the abrupt change of the MR upon cooling through $T_s$, revealed at ambient and finite pressure, together with the drastic change of the pressure dependence of the resistivity, indicate significant changes in the electronic structure associated with $T_s$. This is consistent with recent results of a mobility spectrum analysis \cite{Huynh2014} and angle-resolved photoemission studies \cite{Shimojima2014, Nakayama2014, Tan2013, Watson2015b}. From the latter works evidence for orbital ordering accompanied by a Fermi surface reconstruction have been derived which sets in at $T_s$. This is consistent with thermodynamic data indicating a well-definded second-order phase transition at $T_s$ \cite{Boehmer2013}.

\section{V. Summary}

Detailed studies of the resistivity and magnetoresistivity were performed under hydrostatic pressure on high-quality FeSe single crystals. From these experiments, the pressure dependences of the tetragonal-to-orthorhombic phase transition at $T_s$ and the superconducting transition could be determined with high accuracy. For $T_s$ we find a linear decrease with increasing pressure at a rate -31K/GPa which is accompanied by a linear increase in $T_c$ with +5.8K/GPa. In addition, the pressure studies revealed a normal-state resistivity which is (i) highly pressure sensitive in the tetragonal phase and (ii) pressure independent in orthorhombic state preceding the superconducting transition. More specifically, we observed an in-$T$ linear resistivity over an extended temperature range the slope of which does not change with pressure. Given the positive effect of pressure on both $T_c$ and the strength of antiferromagnetic spin fluctuations \cite{Imai2009} in this compound, we do not see an obvious way to relate the $T$-linear resistivity to a magnetic quantum-critical-point scenario as, e.g., revealed for  BaFe$_2$(As$_{1-x}$P$_x$)$_2$ \cite{Kasahara2010}. Evidence for ($\pi$,0) spin fluctuations have been found in recent neutron scattering studies on FeSe \cite{Wang2015NMR, Rahn2015} even above $T_s$. However, the fact that these are not observed by NMR \cite{Baek2014, Boehmer2015} suggests that they are gapped and do not affect the resistivity at low $T$. These observations add one more peculiarity by which FeSe differs from the 122 compounds. Measurements of the magnetoresistance both at ambient and finite pressure revealed a clear change upon cooling through $T_s$, indicating a change of the electronic structure associated with the structural transition.

\section{Acknowledgements}
We acknowledge fruitful discussions with R. Valent\'{i} and P. Hirschfeld. Work at the Goethe-University Frankfurt was supported by the Deutsche Forschungsgemeinschaft via the Priority Program SPP 1458. 
 




\end{document}